# *West Nile virus outbreak in Italy modelled with the quantum Game of Life*

Andrea Fontana[1,a], Simone Tambascia[1,a], Ciro Di Carluccio[2], Andrea Esposito[1], Bernardo Spagnolo[3,4] & Andrea M. Chiariello[1,b]

[1] *Dipartimento di Fisica, Università degli Studi di Napoli Federico II, and INFN Napoli, Complesso Universitario di Monte Sant'Angelo, 80126 Naples, Italy.*
[2] *Dipartimento di Ingegneria Chimica dei Materiali e della Produzione Industriale, Università degli Studi di Napoli Federico II, and INFN Napoli, Piazzale Vincenzo Tecchio 80, 80125 Naples, Italy.*
[3] *Dipartimento di Fisica e Chimica, Università degli Studi di Palermo, Viale delle Scienze 18, 90128 Palermo, Italy.*
[4] *Stochastic Multistable Systems Laboratory, Lobachevsky University of Nizhny Novgorod, 23 Gagarin Ave., 603950 Nizhniy Novgorod, Russia.*

[a] *Equal contribution authors.*

[b] *Corresponding author:* andreamaria.chiariello@unina.it.

**Abstract**

In the last years, an anomalously high spreading of West Nile virus (WNV) has been observed in Italy, with particularly high peaks of infections in southern Lazio, Campania and Veneto regions. The main disease vector for WNV is represented by *Culex pipiens* mosquitoes, which spread human infections through their bites. Here, we investigate WNV fever epidemic diffusion during summer season 2025 in Italy through a computational approach based on a quantum version of the Game of Life (GOL) cellular automaton model. Specifically, human dynamics evolves according to the GOL rules, while stochastic dynamics of disease vectors, i.e., mosquitoes, as well as their interaction with humans, simultaneously occur.

We show that this model fits the curves of cumulative infected individuals with high accuracy, either at local and average-regional level, with only optimization of mosquito birth and removal rates parameters. Furthermore, leveraging model flexibility, we show that changes in model parameters values elucidate system response to environmental variations. For instance, we quantify, e.g., the impact of mosquito spreading containment measures or sudden mosquito increasing abundance due to climatic and ecological changes.

Overall, we provide a general, quantitative description of WNV infection spreading in Italy which could represent a supportive tool to test different environmental scenarios and could be useful to devise strategies for decision makers to monitor disease vector dynamics and to control consequent virus diffusion.

## Introduction

Epidemic spread of infectious disease is a paradigmatic multiscale phenomenon in which local contacts among susceptible hosts, infected individuals, vectors and environments generate macroscopic health, ecological and economic consequences.

Epidemics are affected by a number of factors ranging from specific geographical features to the impact of climatic changes, such as global warming. In this regard, frequent cases of tropical viruses, such as Zika (ZIKV) [1–5] and Dengue (DENV) [6–9], as well as parasite induced (e.g., by *Plasmodium falciparum*) Malaria [10–12], are beginning to frequently emerge in territories where those infections are traditionally not common, i.e., beyond tropical and sub-tropical environments. Among these, West Nile virus (WNV), which is a enveloped positive single-stranded RNA virus [13–16] and is related to other viral pathogens of the same flaviviruses family [17], e.g., ZIKV, DENV, yellow fever virus (YFV) [18–20] and Rift Valley fever virus (RVFV) [21–24], poses a paradigmatic example since, in recent years, an increasing number of infections has been reported in countries where the virus was previously uncommon, including several Italian regions that now experience recurrent WNV outbreaks during summer seasons.

Transmission of WNV to humans occurs through infected birds or bites from mosquito carrying the virus, mainly belonging to the *Culex pipiens* species, i.e., common northern house mosquitoes [25,26]. This species has adapted well to urban environments [27] and its anomalous diffusion has been recently shown to be influenced by changes in climatic conditions, such as rising temperatures or rainfall and humidity levels [28,29]. After breeding around houses, *Culex pipiens* larvae are produced and thrive in containers, disposed water-holding vessels, water tanks and open sewers [30]. These mosquitoes are massively spread across the Northern Temperate Zone (NTZ), including North American, European, North African and Asian countries [31], posing challenges for disease vector control due to their daytime activity and limited flight range.

Therefore, from a modelling point of view, WNV is an example of vector-borne disease system in which transmission dynamics is linked to ecological and climatic conditions. Similar considerations apply to DENV, ZIKV and YFV infections, whose transmission is shaped by mosquito ecology (particularly from *Aedes aegypti* species, commonly known as Egyptian mosquito), urban habitat suitability, serotype diversity and climate-sensitive environmental conditions [7,9,32]. Analogously, the spread of plant pathogen induced infections, such as the recent *Xylella fastidiosa* infestation [33–35], depends on a complex interplay of vector behavior, host susceptibility, plant phenology, delayed symptom emergence and the spatial organization of agricultural landscapes [36–38]. Different computational tools and effective models have been developed to describe this kind of epidemic spread mediated by disease vectors [39], including stochastic lattice models [40–42], the recently proposed epidemiological renormalization group (eRG) framework applied to DENV spreading [43] and COVID-19 pandemic wave in Europe and US [44–48], and multiscale modelling approaches [49].

Here, we propose a computational model based on the generalized semi-classical Game of Life (gSCGOL), a quantum version of the famous cellular automaton recently applied to successfully describe the 2020 COVID-19 pandemic in the UK [50], to investigate WNV outbreak in Italy during the 2025 summer season, with a particular focus on Campania, Lazio and Veneto regions, where anomalous waves of infection have been reported [26]. The model incorporates, as its main features, human and mosquito dynamics, modelled according to the gSCGOL rules and to a stochastic spreading process, respectively, as well as contact mechanisms between individuals and disease vectors. We find that the model accurately reproduces the temporal evolution of infections at a weekly resolution, either when epidemiological data are aggregated across different territories and at the regional, local scale. Then, once model parameters are fitted, we further discuss some model predictions about the system response under different environmental scenarios, encoded in variations of these key parameters, such as mosquito birth and mortality rates, which in turn may be linked, e.g., to the effects of environmental and climatic changes.

## Results

### ***The generalized semi-classical Game of Life applied to WNV infection spreading***

The Game of Life (GOL) is a cellular automaton model defined on a two-dimensional (2D) lattice, where each cell can exist in one of two possible states: alive (i.e., 1) or dead (i.e., 0). The state of each cell evolves according to deterministic update rules that depend on the number of living cells within its immediate neighbourhood, consisting of the eight surrounding cells [51]. Thanks to its ability to predict complex emergent behaviours from simple local interactions, the model has been widely applied to several fields, including nonlinear dynamics, statistical mechanics and ecology [52–57]. More recently, the GOL framework has been extended to a generalized semi-classical quantum version (gSCGOL) [58–61] in which the binary classical states, 1 and 0, are replaced by pure quantum binary states (qubits), $|1\rangle = \binom{1}{0}$ and $|0\rangle = \binom{0}{1}$, whose temporal evolution is determined through the application of quantum birth, death, and survival operators. The gSCGOL model has been recently employed to investigate the dynamics of biological species under varying environmental conditions [57], as well as the evolution of the COVID-19 pandemic during the 2020 outbreak [50].

In our approach to modelling WNV outbreak, the system consists of two parallel components: a vulnerable human population and a vector-borne transmission system that simulates mosquito spread. The human population evolves according to the gSCGOL rules, independently of the vector dynamics, while interactions between the two systems account for disease transmission. To model human dynamics, we largely follow the framework and rules described in [50,57]. Specifically, human population is defined on a 100×100 2D lattice, where each cell is associated with a normalized superposition $|\psi\rangle$ of the pure qubits states, $|1\rangle$ and $|0\rangle$, weighted with real parameters $a \in [0,1]$ and $b \in [0,1]$ representing the level of being alive (i.e., *i*-th cell liveness) or dead (i.e., *i*-th cell deadness), respectively (formally: $|\psi\rangle = a|1\rangle + b|0\rangle$). Briefly, as described in [50], the evolution of each cell state is defined by the application of a quantum operator $\widehat{G}$ that is a combination of the above-mentioned birth, death and survival operators (Appendix). The form of $\widehat{G}$ depends on the parameters $A_i = \sum_j a_{j,i}$ (referred to as *i*-th cell neighbourhood liveness, Appendix), where the sum runs over the eight *j*-th neighbors of *i*-th cell ($a_{j,i}$ being their liveness), and on V (referred to as vulnerability), which in our simulations is kept fixed to $V = 1$, without loss of generality. All technical details can be found in [50] and are summarized in the Appendix.
From the human system point of view, each cell represents a fraction of individuals with a certain level of mobility: condition $a = 1$ corresponds to the highest mobility while $b = 1$ to the lowest one. The lattice is initialized as a random uniform distribution of normalized a and b values. The vector system, in contrast, stochastically evolves on a lattice of the same size of the human one, where mosquito birth (α) and removal (β) rate parameters are defined and where each cell is a binary variable representing presence or absence of mosquitoes. At each time step, new mosquitoes are randomly placed in the gird, with a probability which depends on α, then, as simulation proceeds, mosquito counts increases.
WNV infection of humans occurs as follow: at each time step, cells with a non-zero mosquito count interact with their homologous cells of the human lattice, which in turn can be infected with a probability p, simulating mosquito bites events. The number $N_{new}$ of new infected individuals is then given by the following expression: $N_{new} = \frac{N_{pop}}{n^2} p = \frac{N_{pop}}{n^2} v_p a$, where $v_p$ is a constant (measuring the viral power of WNV), a is the liveness of the *i*-th cell, $N_{pop}$ is the approximate population size of the considered territory and n is the lattice size (Appendix). Finally, since WNV infection occurs exclusively through mosquito bites, direct human-to-human transmission has not been included in the present modelling. Nevertheless, variants of such simple model can be easily implemented to account for pathogens spreading through interpersonal contacts, as previously done, e.g., for the COVID-19 pandemic case in UK [50].

***WNV outbreak in Italy, data from 2025 summer season***

The numerical data of WNV infected individuals are taken from the Italian Istituto Superiore di Sanità (ISS) surveillance reports [26]. Infection monitoring covers an approximately four-month period corresponding to the 2025 summer season, with epidemiological data collected and updated through weekly bulletins. Panels in Fig. 1a display detailed heatmaps of infected individuals across all Italian provinces during July, August and September 2025 and reveal that territories in Campania, Lazio and Veneto regions exhibit the highest infection levels.

For the modelling purposes, we select 15$^{th}$ July as initial reference date, corresponding to the onset of the rapid growth phase of infections. In particular, Lazio and Campania regions exhibit a steep increase in the number of cases, followed by a plateau in the cumulative number of infected individuals around mid-September, reaching approximately 250 and 120 cases, respectively. In contrast, Veneto region shows a slower growth rate, with plateau (approximately 100 cases) occurring nearly one month later (Fig. 1b).

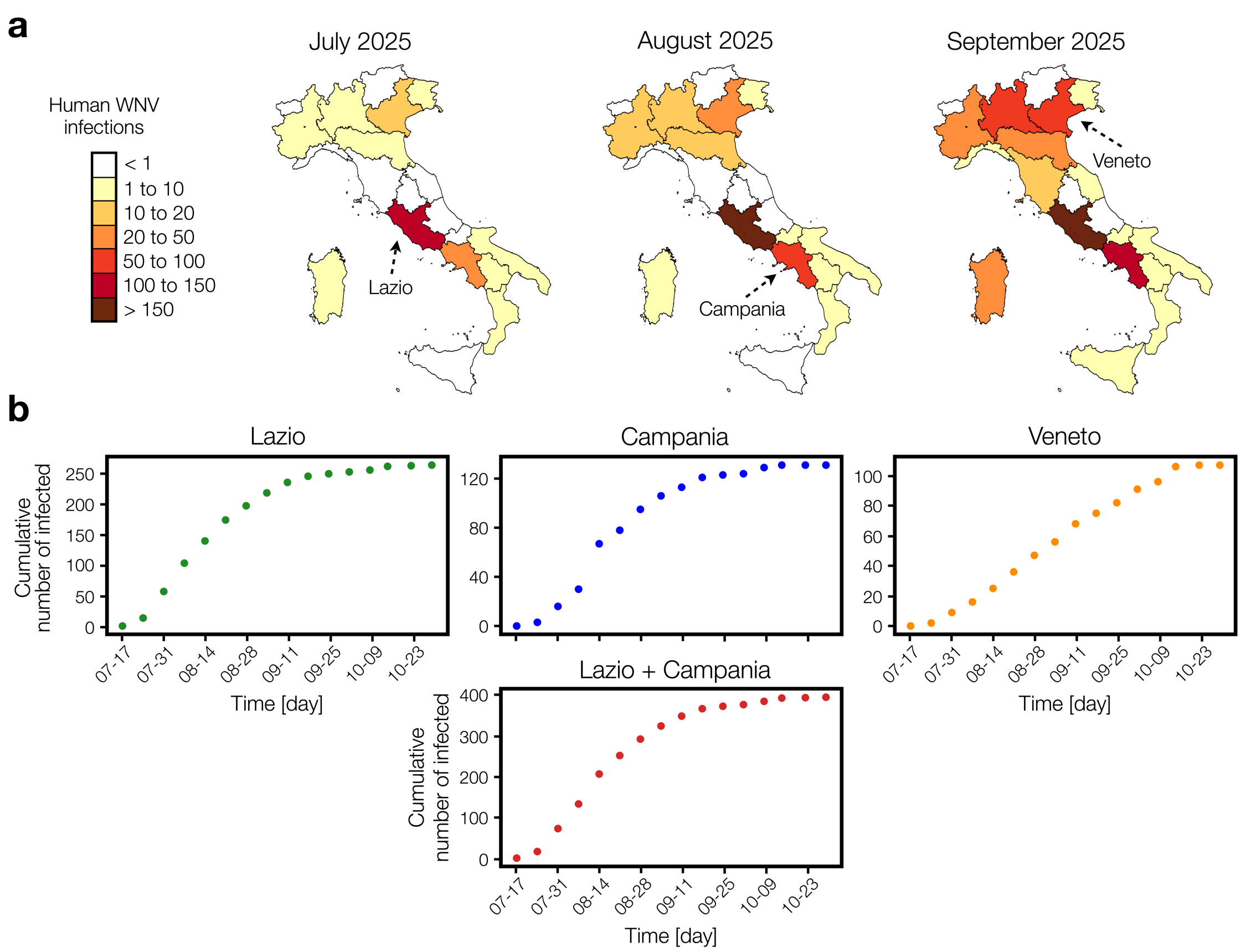

**Figure 1: *West Nile virus outbreak in Italy during summer 2025.*** **a**: Heatmaps illustrating spatial distribution and progression of West Nile virus (WNV) infections across Italian regions in July (left), August (center) and September (right) 2025. Colour scale indicates the number of reported infections in each province. **b**: Weekly time series of reported WNV infection cases in Lazio, Campania and Veneto regions (upper left, central and right panels) and aggregated incidence curve for Lazio and Campania (lower panel) during summer 2025. Data taken from [26].

***Simulation of 2025 WNV outbreak dynamics and model performances***

Simulations of human population dynamics starts from 15th July and are carried out until 1st November, with each simulation step corresponding to one day. Mosquitoes are introduced into the vector lattice from the first simulation day with probability linked to the model parameter α (Fig. 2a), which is assumed to remain constant during the first part of the simulation, mimicking the initial phase of the outbreak, approximately until the curve of cumulative number of infections exhibits a change in concavity, i.e., around mid-August. After this stage, we make the simple yet ecologically plausible ansatz that mosquito population progressively decrease through a gradual removal process, modulated by the parameter β (Figs. 2b,c).

To test the performance of the model, we first considered the aggregated infection dynamics obtained by summing the reported cases from both Lazio and Campania regions. We find that the model is able to reproduce an *in-silico* infection curve, as a function of time, that describes extremely well the observed cumulative number of infected individuals (Fig. 3a). The corresponding optimized parameter values are reported in Table 1. Therefore, these result show that the gSCGOL model is capable of accurately capturing the epidemic time evolution at average level.

Next, we tested whether the model is able to describe epidemic dynamics at regional level. To this aim, we separately analysed the infection curves for Lazio, Campania and Veneto regions, performing independent simulations by appropriately re-scaling model parameters previously optimized for the aggregated infection curve. In all cases, the model is able to accurately capture the temporal evolution of WNV transmission, with Lazio (Fig. 3b) and Campania (Fig. 3c) regions exhibiting pretty similar epidemic dynamics, characterized by an equally good agreement with observed data and comparable optimized parameter values. Conversely, the epidemic curve observed in Veneto region exhibits a less sharp dynamics, characterized by a smoother increase in the number of infected individuals (Fig. 3d). The model is again able to accurately capture this behavior, upon keeping mosquito dynamics stationary for approximately 2-3 weeks after 1st September, reflecting region-specific spreading mechanisms likely linked to geographical and climatic differences.

The optimized values of model parameters α and β, with respect to the observed data, for the three considered regions are reported in Table 1. Taken together, those results highlight that the gSCGOL approach provides a valuable tool for accurately modelling real WNV epidemiological data at both global and regional scales.

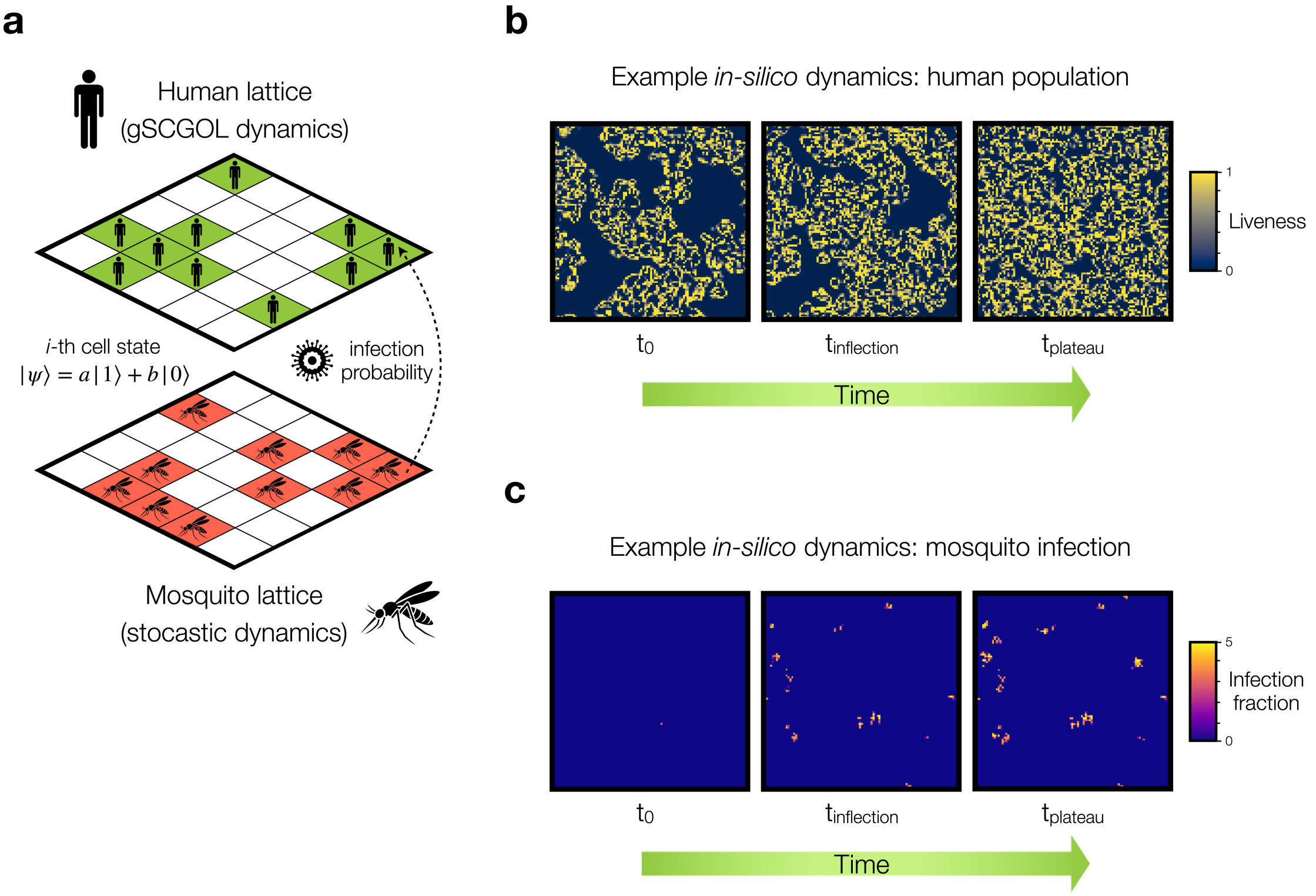

**Figure 2:** ***Computational model of human population and disease vectors dynamics*****.** **a**: Human individuals evolve on a two-dimensional (2D) lattice according to the generalized semi-classical Game of Life (gSCGOL) dynamics rules, encoded in specific quantum operators (upper panel). At the same time, mosquitoes evolve stochastically on a homologous 2D lattice (bottom panel) and interact with human individuals, at any time step, with a fixed probability. **b**: Examples of simulated configurations of the human population at different time stages of the epidemic trajectory: initial phase (left), inflection point of the infection curve (center) and plateau regime (right). Each pixel represents corresponding liveness level of individuals. **c**: As **b**, for the infected individuals. Each pixel represents a fraction of infected individuals.

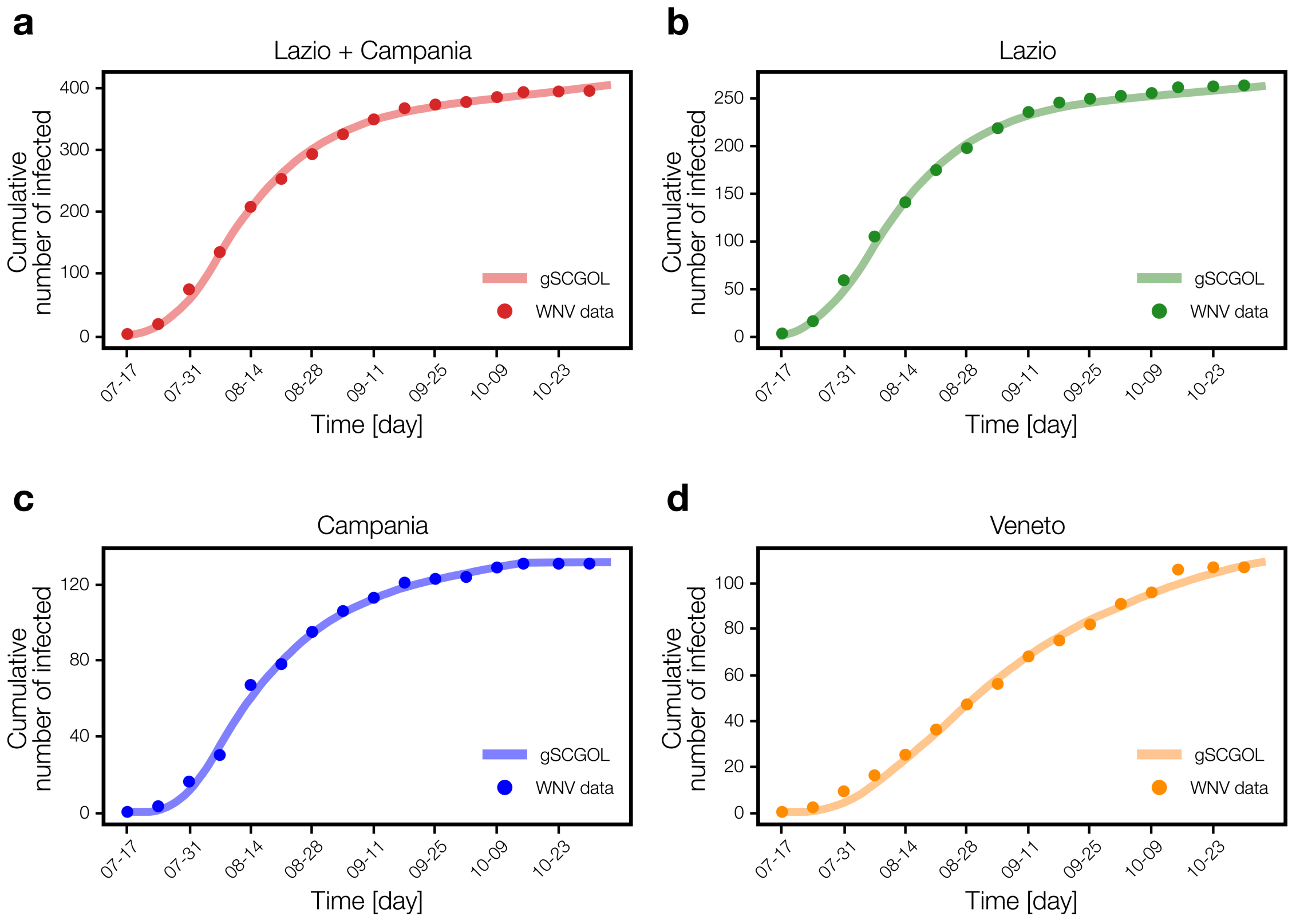

**Figure 3:** ***Model performances in describing aggregate and regional infection dynamics in summer 2025.*** **a**: Simulated epidemic curve generated by the gSCGOL model (line) compared with aggregated number of reported infected individuals data (dots) over Lazio and Campania regions. **b**-**d**: As **a**, for the individual regions of Lazio (**b**), Campania (**c**) and Veneto (**d**). Simulated trajectories represent averages over 100 independent runs, performed using optimal parameters (data taken from [26]).

| | ***Best α*** | ***Best β*** |
|---|---|---|
| ***Lazio + Campania*** | 1.0 | 0.046 |
| ***Lazio*** | 1.0 | 0.048 |
| ***Campania*** | 1.0 | 0.043 |
| ***Veneto*** | 0.4 | 0.012 |

**Table 1:** Optimized values of model parameters α and β, with respect to observed data, for the cumulative simulation across Lazio and Campania regions, and for the simulations performed separately for each individual region (Lazio, Campania and Veneto).

***Modelling of other WNV infection waves and the effect of environmental factors***

A growing amount of evidence is highlighting the triggering role of climate change and global warming in driving the anomalous proliferation of species that may act as disease vectors.

Motivated by these considerations, we analyzed previous WNV outbreaks in Italy, focusing on the entire Italian territory in 2018, 2022 and 2025, where relevant epidemic waves have been recorded [26]. The gSCGOL model confirms to be an excellent tool for describing epidemic spreading even in those cases (Figs. 4a-c), as shown by the good agreement between model predictions and the cumulative number of infected individuals [26]. Interestingly, the data reveal an increasing trend in the number of infections over the last eight years, which is reflected in the progressively larger optimal values of the parameter α, estimated from the model calibration.

In order to find a possible connection between the parameter α and environmental conditions, we correlated its optimized values with temperature and relative humidity levels measured in 2018, 2022, and 2025, by using data from Copernicus Earth observation component of the EU’s Space programme [62]. Since infection curves start from mid-July, we considered monthly average climatic values of June, under the assumption that environmental conditions affect mosquito abundance with a delay of approximately 2-5 weeks, in line also with recent reports [28,29,63], and geographical averages computed over territories involving Veneto, Emilia-Romagna, Lazio and Campania regions, where most of the infections were recorded. Intriguingly, the parameter α exhibits a trend consistent with the concomitant increase in average temperature (Fig. 4d), although only few points do not allow to derive a reliable correlation statistic. Conversely, the parameter α appears to be negatively correlated with relative humidity, which shows a decreasing trend during the 2018-2025 period (Fig. 4e), in agreement with mosquito monitoring reports [28,29,63].

The same analysis, restricted only to Veneto and Emilia-Romagna regions, where most infections were concentrated in 2018 and 2022, but dropped in 2025, did not reveal any obvious correlation between environmental features and model parameters, suggesting that additional factors, e.g., rainfall patterns, should be taken into account to establish connections between model parameters and local climatic conditions [64–66]. Taken together, these findings suggest an interplay between temperature, relative humidity and infection dynamics, similarly to what has been reported for DENV transmission [43], and, within the gSCGOL framework, these environmental effects are encoded through variations in the mosquito growth parameter, consistently with previous ecological and epidemiological studies [28,29,63–65].

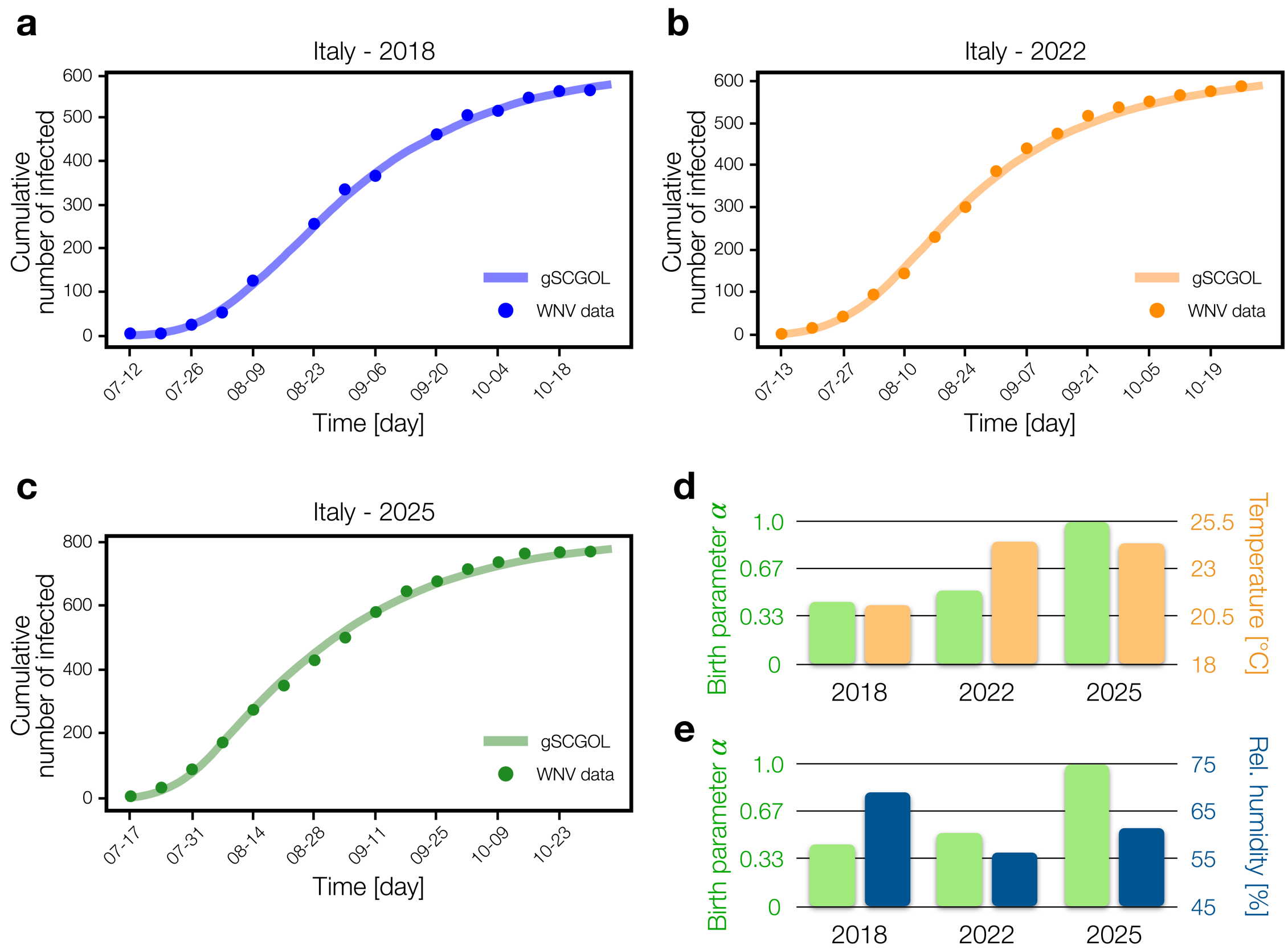

**Figure 4:** ***Comparative modelling of recent WNV infection waves and their dependence on climatic changes.*** **a-c**: Simulated epidemic curves of WNV infection waves during 2018 (**a**), 2022 (**b**) and 2025 (**c**) summer seasons (lines), compared with the corresponding total number of reported infections (dots) in Italy (infection data taken from [26]). **d**: Optimal values of the model parameter α, obtained from model fitting in different years (i.e., 2018, 2022 and 2025), compared with the corresponding monthly average temperature (taking June as reference month). **e**: As **d**, compared with corresponding monthly average relative humidity (taking June as reference month). Climatic data taken from [62].

### *Stability of the system and model predictions*

Next, we investigate the stability of the model with respect to variations of system parameters values. Once calibrated, the gSCGOL model provides a computational tool to explore how changes in model parameters may reflect different ecological and environmental scenarios. For instance, mosquito birth and removal rates can be associated with climatic variables such as temperature, rainfall and relative humidity levels. This approach is therefore useful for quantitatively assessing the environmental drivers underlying vector proliferation and disease spreading, as well as for evaluating system's response to prevention strategies, including mosquito population control measures.

We first considered parameter variations leading to an increase in the mosquito population. Within the model, this can occur through two distinct mechanisms: an increase in the mosquito birth rate α or a decrease in the mosquito removal rate β. As representative examples, we performed simulations of the system with a 10% (Fig. 5a) and 50% (Fig. 5b) increase (decrease) of the birth (removal) rate with respect to the values optimized from the aggregated (across Lazio and Campania regions) 2025 epidemic data.

Analogously, we investigated parameter variations associated with a reduction in the mosquito population, thereby simulating, e.g., the potential effects of hypothetical vector-control interventions implemented by local administrations. As before, within the model, this is easily achieved by decreasing (increasing) mosquito birth (removal) rates. Again, we performed simulations considering 10% (Fig. 5c) and 50% (Fig. 5d) re-modulations in α and β model parameters. Together with the previously discussed dependence of model parameters on environmental variables, these simulated epidemic curves provide a framework for exploring possible outcomes of epidemic evolution predictions and assessing the effectiveness of control strategies.

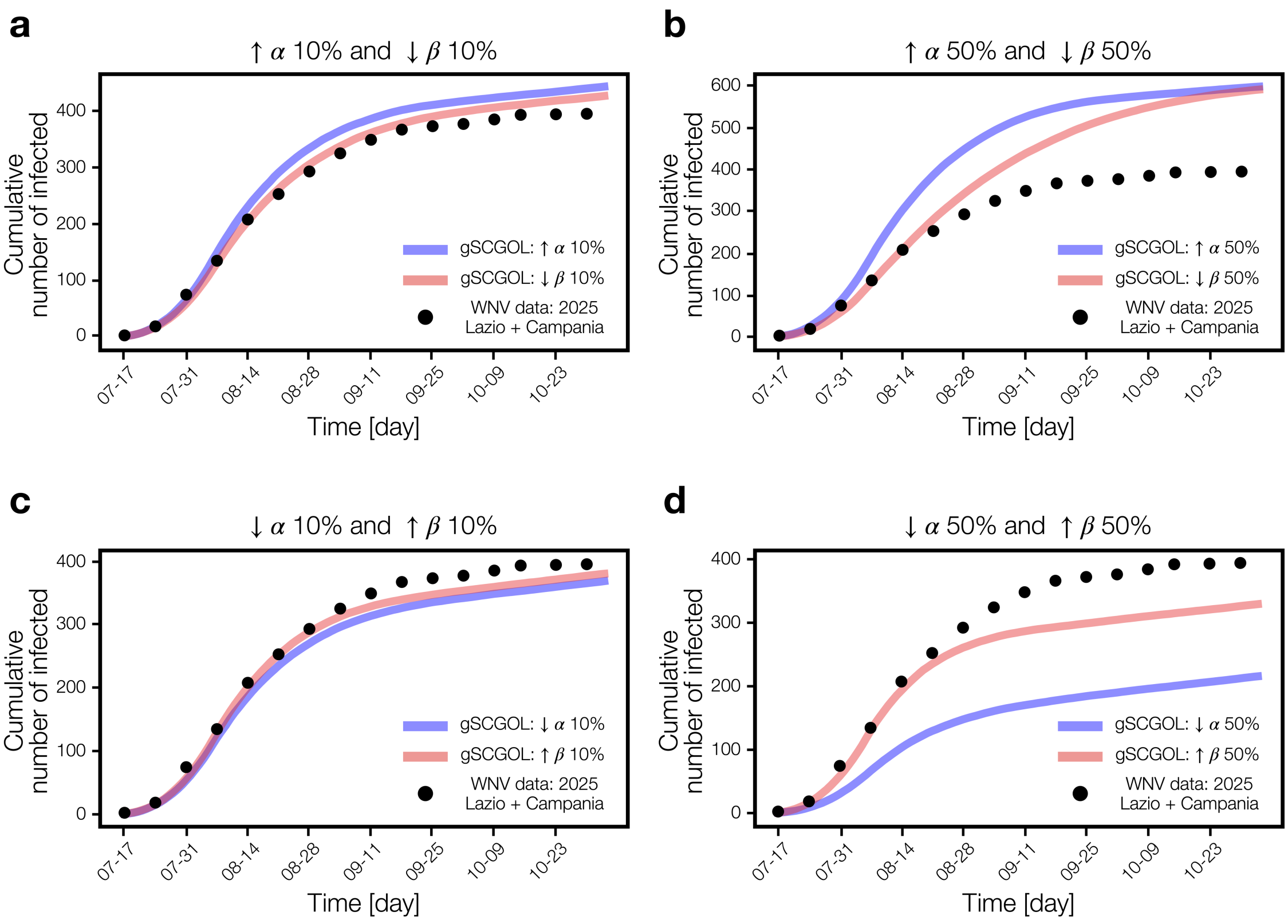

**Figure 5:** ***Model stability upon variation of system parameters and predictions*****.** **a**: Model response to a 10% increase in the parameter α with respect to its optimal value (blue curve) and to a 10% decrease in the parameter β with respect to its optimal value (red curve) from aggregated (across Lazio and Campania regions) 2025 epidemic data (black dots) [26]. **b**: As **a**, for a 50% increase (decrease) in α (β) parameter. **c**: Model response to a 10% decrease in the parameter α with respect to its optimal value (blue curve) and to a 10% increase in the parameter β with respect to its optimal value (red curve) from aggregated (across Lazio and Campania regions) 2025 epidemic data (black dots) [26]. **d**: As **c**, for a 50% decrease (increase) of α (β) parameter.

## Conclusions

In this work, we investigated the temporal evolution of the West Nile virus (WNV) epidemic in Italy during the 2025 summer season [26]. To quantitatively describe epidemiological data of infected individuals, we employed the generalized semi-classical Game of Life (gSCGOL) model [59], the quantum version of the famous cellular automaton, which has recently been applied to the study of COVID-19 pandemic dynamics [50]. We found that the gSCGOL framework provides an effective and accurate description of WNV spreading across different Italian regions during the 2025 outbreak. Furthermore, we also showed that the model is able to accurately recapitulate epidemic waves observed in previous years and reveals an increasing national trend in WNV infections between 2018 and 2025.

Within the model, this increasing trend is encoded in the parameter governing mosquito increase rate, whose optimized values have been compared with variations in summer temperature and relative humidity levels measured in those regions accounting for the majority of reported infections [62]. Such comparison suggests a potential link between mosquito proliferation, increasing temperatures and decreasing relative humidity levels [28,29,63,64], although a larger number of temporal observations would be necessary to establish statistically robust correlations. In this regard, an analysis of model response to change of system parameters has been performed. Further developments of such framework could incorporate additional environmental features, such as rainfall levels, directly into model parameters in order to enhance the interpretability of results and the predictive capability of the approach, especially at local geographical scales.

Overall, our results show that the gSCGOL model can be employed to effectively describe the spreading dynamics of WNV and that, thanks to its generalizability, it could in principle be extended to other arboviral diseases and vector-borne pathogens worldwide, such as DENV, ZIKV and YFV, as well as to different ecological systems, e.g., the ongoing proliferation of *Karenia* planktonic organism in marine environments of Southern Australia [67,68]. More generally, these findings highlight the importance of developing computational models capable of describing relevant ecological and epidemiological processes, with few parameters encoding key dynamical features of the system.

## Appendix

In our simulations, we largely employ generalized semi-classical Game of Life (gSCGOL) rules described in [50]. The states of cells on the human population lattice, represented by pure quantum binary states $|1\rangle = \begin{pmatrix}1\\0\end{pmatrix}$ and $|0\rangle = \begin{pmatrix}0\\1\end{pmatrix}$, evolve in time under the action of the operator $\widehat{G}$, which is a linear combination of the basic quantum birth ($\widehat{B}$), death ($\widehat{D}$) and survival ($\widehat{S}$) operators, defined as follows:

$$\widehat{B} = \begin{pmatrix}1 & 1\\0 & 0\end{pmatrix},\ \widehat{D} = \begin{pmatrix}0 & 0\\1 & 1\end{pmatrix},\ \widehat{S} = \begin{pmatrix}1 & 0\\0 & 1\end{pmatrix}.$$

The specific form of the operator $\widehat{G}$ depends on the $i$-th cell liveness $A_i = \sum_j a_{j,\,i}$, according to the following rules, described in [59]:

$$\begin{aligned}
\widehat{G} &= \widehat{D} && \text{if} \quad A_i < A_0 - 2V \ \text{or} \ A_i > A_0 + V\\
\widehat{G} &= \left(1+\sqrt{2}\right)(A_0 - V - A_i)\widehat{D} + [A_i - (A_0 - 2V)]\widehat{S} && \text{if} \quad A_0 - 2V < A_i < A_0 - V\\
\widehat{G} &= \left(1+\sqrt{2}\right)(A_0 - A_i)\widehat{S} + [A_i - (A_0 - V)]\widehat{B} && \text{if} \quad A_0 - V < A_i < A_0\\
\widehat{G} &= \left(1+\sqrt{2}\right)(A_0 + V - A_i)\widehat{B} + (A_i - A_0)\widehat{D} && \text{if} \quad A_0 < A_i < A_0 + V
\end{aligned}$$

where $A_0$ defines the vulnerable population density and is fixed at $A_0 = 3$ in all the simulations, without loss of generality, while the vulnerability parameter V determines the threshold changes in the $i$-th cell Moore neighbourhood liveness $A_i$ at which the operator changes occur and is fixed at $V = 1$, also without loss of generality. These choices are consistent with the parameterization introduced in [59]. Allowing V to vary during simulations would provide a means of modelling containment interventions including, e.g., population lockdowns and restrictions on mobility.

The mosquito population increases by one cell per day with probability α and decreases, after the inflection point of the epidemic curve, at a rate β. These two quantities are the parameters optimized to obtain the best fit to the infection curves across the different datasets. The remaining parameter values used for the infection dynamics are as follows: viral power is set to $v_p = 0.002$; approximate population sizes are $N_{pop}^{Lazio} = 5710000$, $N_{pop}^{Campania} = 5590000$, $N_{pop}^{Veneto} = 4850000$ for regional simulations (Figs. 2, 3 and 5) and $N_{pop}^{Italy} = 59000000$ for national simulations (Fig. 4); lattice size n is set to n = 100 for regional simulations (Figs. 2, 3 and 5) and to n = 200 for national simulations (Fig. 4).

Raw dew point temperature data [62] were converted to relative humidity by means of the August-Roche-Magnus formula [69]. Specifically, for each observation, the actual vapour pressure $e_s(T_d)$ is estimated from the dew point temperature $T_d$ (given in °C), while the saturation vapour pressure $e_s(T)$ from the corresponding temperature T (given in °C) [62]. Relative humidity was then obtained as the percentage of ratio of these two quantities: $RH = 100 \cdot \frac{e_s(T_d)}{e_s(T)}$, where $e_s(T) \propto \exp\left(\frac{17.625 \cdot T}{243.04 + T}\right)$.

**Competing interests**

The authors declare no competing interest.

**Funding**

No relevant funding.

**Author contributions**

A.M.C. conceived and supervised the project. S.T. developed the codes. A.F. and S.T. run computer simulations. C.D.C. analyzed and processed temperature and humidity data. A.M.C., A.F. and B.S. made figures assembling and wrote the paper with input from other authors.